\documentclass[cup6b,epsf]{cupbook}

\usepackage{epsf}

\newcommand{\msun}{\mbox{${M}_{\odot}$}}

\begin{document}

\pagenumbering{arabic}


\author{Vikram Dwarkadas\\ ASCI FLASH Center, Dept.~of Astronomy and
Astrophysics, University of Chicago,\\ 5640 S Ellis Ave, Chicago IL
60637}

\chapter{The Evolution of Supernovae in the Winds of Massive Stars}

\begin{abstract}
We study the evolution of supernova remnants in the circumstellar
medium formed by mass loss from the progenitor star. The properties of
this interaction are investigated, and the specific case of a 35
$\msun$ star is studied in detail. The evolution of the SN shock wave
in this case may have a bearing on other SNRs evolving in wind-blown
bubbles, especially SN 1987A.
\end{abstract}

\section{INTRODUCTION}
Type II Supernovae are the remnants of massive stars (M $>$ 8
M$_{\odot}$). As these stars evolve along the main sequence, they lose
a considerable amount of mass, mainly in the form of stellar
winds. The properties of this mass loss may vary considerably among
different evolutionary stages. The net result of the expelled mass is
the formation of  circumstellar wind-blown cavities, or bubbles,
around the star, bordered by a dense shell. When the star ends its
life as a supernova, the resulting shock wave will interact with this
circumstellar bubble rather than with the interstellar medium. The
evolution of the shock wave, and that of the resulting supernova
remnant (SNR), will be different from that in a constant density
ambient medium.

In this work we study the evolution of supernova remnants in
circumstellar wind-blown bubbles. The evolution depends primarily on a
single parameter, the ratio of the mass of the shell to that of the
ejected material. Various values of this parameter are explored. We
then focus on a specific simulation of the medium around a 35 $\msun$
star, and show how pressure variations within the bubble can cause the
shock wave to be corrugated. Different parts of the shock wave collide
with the dense shell at different times. Such a situation is
reminiscent of the evolution of the shock wave around SN 1987A.

\section{The SN Profile}

The interaction of the SN shock wave with the surrounding medium in
the early stages depends on the density profile of the SN and the
surrounding medium. The density structure of the ejecta depends on the
structure of the star and the shock acceleration in the outer layers
(Chevalier \& Fransson 1994). Although observational information on
valid density profiles is scarce, numerical simulations, especially
for SN 1987A, as well as semi-analytic calculations show that the
ejecta density profile in the outer layers can be approximated by a
power-law in radius (Fig 1.1; see Chevalier \& Fransson 2003 for further
information). This approximation is used in the current work.

\begin{figure}
\begin{center}
\leavevmode\epsfxsize=8cm \epsfbox{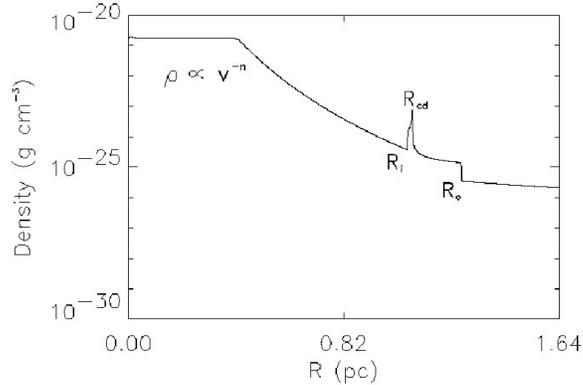}
\end{center}
\caption{The density profile of the ejected SN material in the initial
  stages, from a numerical calculation. Interaction of the ejecta with
  the ambient medium gives rise to an outer shock (R$_o$), inner shock
  (R$_i$) and contact discontinuity (R$_{cd}$).  }
\label{fig:snprof}
\end{figure}

\section{Structure of the Circumstellar Medium (CSM)}

The general structure of a wind-blown nebula was first elucidated by
Weaver et al.~(1977). In the simplest, two-wind approximation, a fast
wind from a star collides with slower material emitted during a
previous epoch, driving a shock into the ambient medium. The pressure
of the post-shock material causes the freely flowing fast wind to
decelerate, driving a second shock that travels inwards towards the
center. A double-shocked structure, separated by a contact
discontinuity, is formed. Figure 1.2 shows the density and pressure
profiles from a simulation of a wind-blown bubble. Proceeding in the
direction of increasing radius from the central star we find the
following regions delineated: freely flowing fast wind, inner or
wind-termination shock (R$_{\rm t}$), shocked fast wind, contact
discontinuity (R$_{\rm cd}$), shocked ambient medium, outer shock
(R$_{\rm o}$) and unshocked ambient medium.

\begin{figure}
\begin{center}
\leavevmode\epsfxsize=10cm \epsfbox{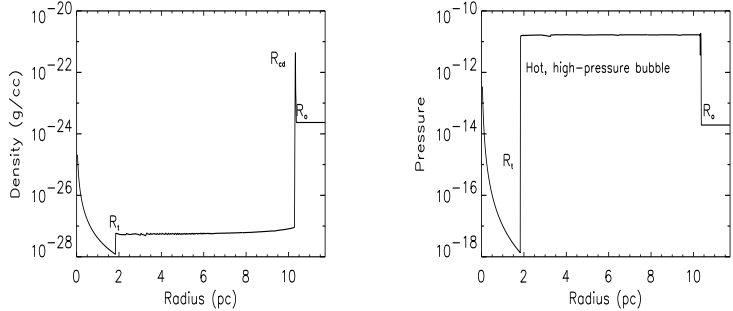}
\end{center}
\caption{a) Density and b) Pressure profiles from a numerical
simulation of a wind-blown bubble around a massive star. }
\end{figure}

\section{SNR-Circumstellar Medium Interaction}

The interaction of the SN ejecta with the freely expanding wind gives
rise to a double-shocked structure, consisting of a forward shock
driven into the wind and a reverse shock moving into the ejecta. Given
the low density interior of the wind-blown bubble, the luminosity of
the remnant is lower than if the explosion were to occur within the
ISM. It is clear that in general most of the bubble mass is contained
within the dense circumstellar (CS) shell. Thus the interaction of the
ejecta with this shell is crucial to determining the evolution of the
remnant. This interaction depends on a single parameter $\Lambda =
\frac{M_{shell}}{M_{ejecta}}$, the ratio of the shell mass to the
ejecta mass.

An exploration of the interaction of SN shock waves with CS bubbles
described by the Weaver et al.~(1977) model shows (eg.~Dwarkadas
2002), that for small values of the parameter $\Lambda \le 1$, the
structure of the density profile is important (Figure 1.3a-d). Just
after the shock-shell interaction has taken place, the density {\it
decreases} outwards from the reflected shock to the contact
discontinuity. However as the evolution proceeds, the supernova
remnant begins to ``forget'' the existence of the shell, and loses
memory of the interaction. The density structure changes to reflect
this, and begins to {\it increase} from the reflected shock to the
contact discontinuity.  In this case it takes about 15 doubling times
of the radius for the remnant to forget the interaction with the shell
(Fig 1.3d). In another few doubling times, the remnant density profile
will resemble that of a SNR evolving directly in the ambient
medium. When computing the X-ray or optical emission from the remnant,
which are functions of the density of the shocked material, it is
imperative that this changing density structure be taken into account.

As the value of the parameter $\Lambda$ increases, i.e. the mass of
the wind-blown shell increases compared to the ejecta mass, the energy
transmitted by the remnant to the shell also increases. Energy
transfer to the shell becomes dynamically important, and the remnant
evolution is speeded up. The reflected shock moves rapidly through the
ejecta, and complete thermalization of the ejecta is achieved in a
shorter time as compared to the SN reverse shock thermalizing the
ejecta. If the value of $\Lambda$ is large, the SN shock may become
radiative, and the kinetic energy is converted to thermal energy. In
extreme cases, the remnant may then go directly from the
free-expansion stage to the radiative stage, by-passing the classical
adiabatic or ``Sedov'' stage.

\begin{figure}
\begin{center}
\leavevmode\epsfxsize=10cm \epsfbox{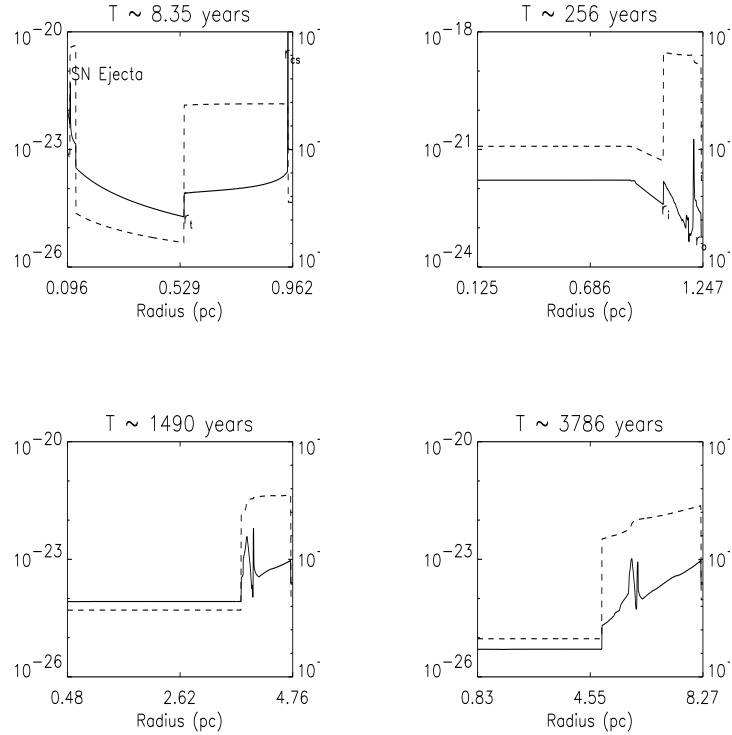}
\end{center}
\caption{Snapshots in time from a simulation of SN ejecta interacting
with a CS bubble. The mass in the circumstellar shell is 14\% of that
in the ejecta. The solid lines display density, with the scale given
on the LHS. The dashed lines displays the pressure, with the scale
given on the RHS. All units are CGS. The time is given at the top of
each figure in years. The labels (a) to (d) in the text go in order of
increasing time, from top to bottom and left to right. }

\end{figure}

\section{A 35 M$_{\odot}$ Star}

Using mass-loss data kindly provided to us by Norbert Langer, we have
 modeled the evolution of the medium around a 35 $\msun$ star, and
 the further interaction of the shock wave with this medium once the
 star explodes as a SN. The star goes through the sequence O-Star,
 Red-Supergiant Star (RSG) and Wolf-Rayet (WR) star. Below we
 describe, mainly through images of the fluid density, the subsequent
 evolution of the CSM around the star.

{\sc \underline{Main Sequence (MS) Stage}}

\begin{figure}
\leavevmode\epsfxsize=10cm \epsfbox {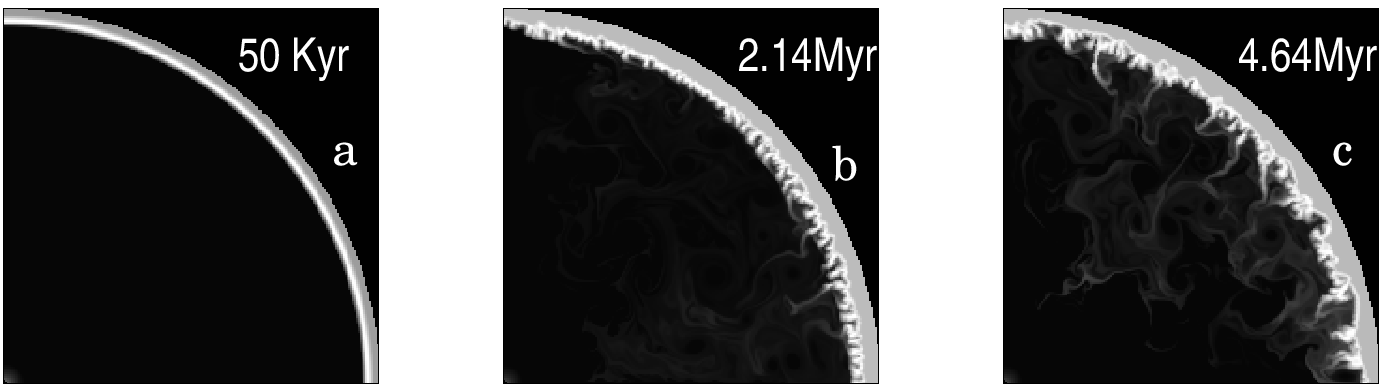}
\caption{Time-sequence of images of the formation of the CSM around a
35 $\msun$ O-Star during the main sequence.  }
\end{figure}

The wind from the star, with velocity of a few (3-4) thousand km/s and
mass loss rate on the order of 10$^{-7} \msun$/yr expands into a
medium with density of about 1 particle/cc, giving rise to a bubble
about 74 pc in radius. Fig 1.4 shows (from left to right) a
time-sequence of density images of the formation of the MS
bubble. Note that the main-sequence shell is unstable to a
Vishniac-type thin-shell instability. The density inhomogeneities lead
to pressure fluctuations which propagate within the interior, which
soon develops into a turbulent state. The evolution of these
perturbations distinguishes our results from those of Garcia-Segura et
al.~(1996), who considered the MS shell to be stable and therefore
assumed spherical symmetry. However the 2D structure is quite
different, and has significant implications for the succeeding
evolution of the bubble.

{\sc \underline{Red-Supergiant (RSG) Stage}}

In the RSG stage the wind velocity falls to a low value of about 75
km/s, whereas the mass loss rate jumps up to a few times 10$^{-5}
\msun$/yr. A new pressure equilibrium is established, and a RSG shell
is formed in the interior, which is also unstable to thin-shell
instabilities. Fig 1.5 (frames 1 and 2) shows images of the density
during the RSG evolution.

\begin{figure}[htbp]
\leavevmode\epsfxsize=10cm \epsfbox{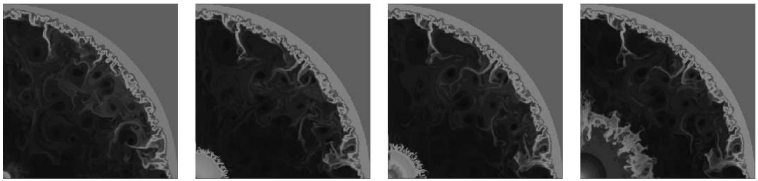}
\caption{The first two density images from left show the formation of
  the inner RSG shell, which is unstable to thin-shell
  perturbations.The next two display the onset of the WR wind and its
  collision with the RSG shell, causing it to fragment and the RSG
  material to be mixed in with the rest of the nebula. }
\end{figure}

{\sc \underline{Wolf-Rayet (WR) Phase}}

The wind velocity in the WR phase climbs back up to almost 3000 km/s,
whereas the mass loss rate drops by only a factor of a few from the
RSG stage. The momentum of the WR wind is then about an order of
magnitude larger than that of the RSG wind, and the wind pushes the
RSG shell outwards, simultaneously causing it to fragment (Fig 1.5,
frame 3). The RSG wind material is mixed in with the rest of the MS
material (Fig 1.5, frame 4), a key result since the RSG wind velocity
was so low that the material by itself could not have gone very
far. Out of $\sim 26 \msun$ of material lost in the wind, about 19
$\msun$ is lost in the RSG stage, so much of the material within the
nebula is composed of matter lost in the RSG phase.

{\sc \underline{SN-CSM Interaction}}

At the end of the WR phase the stellar mass is about 9.1 $\msun$. We
assume that about 1.4 $\msun$ remains as a neutron star, and the
remaining mass is ejected in a supernova explosion, with a density
profile that goes as a power-law in the outer parts, with a power-law
index of 7. The interaction soon forms the usual double-shocked
structure. In Figure 1.6 we show images of the fluid pressure. This
variable is chosen to clearly illustrate the shocked region between
the inner and outer shocks. The shock starts off as a spherical shock
(Fig 1.6a), but the pressure within the turbulent interior soon causes
it to become rippled (Fig 1.6b). The corrugated shock structure
collides with the boundary of the bubble in a piecemeal fashion (Fig
1.6c), and as each small part collides with the outer boundary, a
reflected shock arises in that region. There exist many pieces of
reflected shock that arise from various interactions, have different
velocities, and consequently reach the inner boundary at different
times. The thermalization of the material then occurs in different
stages, and X-ray images will reveal a very complicated structure
which will differ considerably on scales of tens to hundreds of years.

HST images of SN 1987A have revealed the presence of various bright
spots around the circumstellar ring, presumably due to the interaction
of the SN shock front with the equatorial ring structure (eg.~Sugerman
et al. 2002). The collision of a highly wrinkled shock with various
parts of the circumstellar shell, leading to the different parts
brightening up at different times, is very similar to the current
situation of the shock front in SN 1987A. The case of SN 1987A however
is more complicated in that the region interior to the ring is
presumed to be an ionized HII region (Chevalier \& Dwarkadas 1995). It
is possible though that an aspherical HII region would serve only to
accentuate the asphericity in the shock front. The simulation
described herein is for a 35 $\msun$ star, whereas in 87A the
progenitor star was less massive, and possibly part of a binary
system. Nevertheless the similarities are striking, and suggest the
existence of such wrinkled shock fronts when SNe evolve in wind-blown
bubbles.

\begin{figure}[htbp]
\leavevmode\epsfxsize=10cm \epsfbox{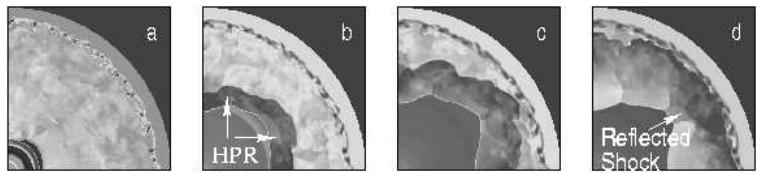}
\caption{Time-sequence of pressure images of the interaction of the
 SNR shock with the WR bubble. HPR represents the high-pressure region
 between the inner and outer SNR shocks. Note the rippled structure of
 the outer shock from frame 2 onwards, and its interaction at
 different times with various parts of the shell.}
\end{figure}

{\it Acknowledgements }

This research is supported by Award \# AST-0319261 from the National
Science Foundation. Vikram Dwarkadas is also supported by the
US.~Department of Energy grant \# B341495 to the ASCI Flash Center (U
Chicago). I would like to thanks the organizers for inviting me to a
most stimulating conference, and to wish Craig Wheeler a very happy
60th birthday.

\end{document}